# Preventing wireless deauthentication attacks over 802.11 Networks

Ananay Arora
<ananay@asu.edu>

## Abstract

The objective of this paper is to investigate a special Denial of Service (DoS) attack against 802.11 wireless networks. This attack is known as the Deauthentication / Disassociation attack which is launched against 802.11-based wireless networks. When a client needs to disconnect from the wireless access point, it sends special frames known as deauthentication or disassociation frames. Due to being encrypted, these frames do not require an authenticated user. Hence, an attacker can craft these frames and send them to the access point in such a way that the access point assumes the frames to be coming from the client and not the attacker. In this paper, an efficient solution is proposed to prevent deauthentication attack by using a session management system used to then verify deauthentication frames.

## 1. Introduction

With an extremely large number of devices like smartphones, laptops, tablets, smartwatches, IoT devices et al, the number of 802.11 networks has increased by a huge amount over the past few years. However, these devices have become easy targets of attack as attackers can easily target devices in wireless proximity and don't require physical interaction with any of the devices. One type of attack, known as the deauthentication / disassociation attack, is described as follows.

To be able to connect to an access point (AP), the client needs to associate with the AP before it can begin exchanging data messages. Prior to association, the client needs to complete the authentication procedure. However, in case the client wants to disconnect, it needs to send a disassociation frame to the AP. Alternatively, in case of a client unexpectedly leaving an AP, it sends a deauthentication frame.

According to the 802.11 network standards, the deauthentication or disassociation frames are unencrypted and do not require authentication. Due to this, an attacker can easily spoof the MAC Address of the Client or the AP in order to make deauthentication requests on behalf of the Client or the Access Point. In order to distinguish between legitimate and non-legitimate deauthentication frames, we must verify the source of these frames. In this paper, a solution is proposed and implemented to verify the source of deauthentication frames sent over 802.11 networks.

## 2. Deauthentication Attacks and Related Work

**2.1 Deauthentication Attack**

The 802.11 Association process can be explained as going through the following states[3]:

1. Unauthenticated and Unassociated
2. Authenticated but Unassociated
3. Authenticated and Associated
4. Authenticated and Associated and 802.1x Authenticated

Both the client and the AP start with state 1. Upon wanting to join a network, the client begins scanning all the channels for a specific AP. When the AP is found broadcasting on a particular channel, the client and the AP perform authentication by exchanging messages. APs can use either an Open Authentication System or one of the following Wireless Security Protocols – WEP (Wired Equivalent Privacy), WPA (Wireless Protected Access) or WPA2 (Wireless Protected Access). In an open authentication system, the AP authenticates any client that tries to join the network. In the case of using a Wireless Security Protocol, the client and AP go through a sequence of challenges and responses using either WEP, WPA or WPA2 encryption protocols. On completion of the authentication procedure, the AP and the client transit to state 2.

In state 2, the client and the AP associate and advance to state 3. In state 3, the Client and the AP start exchanging data packets. If the 802.1x protocol is supported, then the 802.1x authentication messages will be exchanged between the client and the AP. Once the 802.1x authentication is successful, the client and the AP both move to state 4. At any point, on receiving a disassociation message both the client and the AP move back to state 2. Similarly, on receiving a deauthentication message at any point, both the client and the AP move to stage 1. [1]

An attacker can send spoofed deauthentication packets to either the AP or the client and both move to state 1.[1] Tools such as aireplay-ng can be used to send deauthentication packets by spoofing the MAC Address of the attacker to that of any device. In this paper, we present a new protocol based on two-way key and token exchange.

## 2.2 Related Work

Many solutions have been proposed to defend against deauthentication attacks:

**1) Deauthentication using on the letter-enveloped by making use of the factorization problem**

Issues: Generating new primes for every connection is time consuming. This can be DoS Primitive if an attacker keeps sending association requests repeatedly with different prime numbers and different MAC Addresses.

**2) Using R-ARP [6]**
Issues: The client's IP Address can be spoofed by attackers to break the Reverse Address Resolution Protocol. [3][7][1]

**3) Eliminating Deauthentication or Disassociation frames or enqueuing frames for a fixed interval [15]**

> Issues: If an STA associates with multiple APs concurrently, it may cause routing/handoff problems [3][7][1]

**4) Spoofed frame detection based on Frame Sequence Number** [9][10][11][12]
Issues: These frames can be sniffed by the attacker and the sequence number of the next frame can be predicted, if the sequence numbers are assigned deterministically.[3][7][1]

**5) Using 1 bit authentication for management frames** [13][14]
Issues: High probability (50% chance) of guessing the authentication bit [3][7][1]

**6) Modify the current authentication framework to authenticate deauthentication frames**
Issues: Lack of cryptographic primitives in legacy devices [8].

We develop a special method of verifying management frames. We use a one way hashing function, thus making it computationally infeasible to break. This method requires no special cryptographic primitives and thus can be deployed as a simple firmware update.

# 3. Proposed Solution

This research paper proposes an alternative solution to verify the origin of deauthentication frames sent wirelessly over 802.11 networks. The solution makes use of a secure hashing algorithm (SHA) to hash a unique ID during association and checks for the Unique ID when the deauthentication frame is sent.

### 3.1 Universally Unique Identifier (UUID)

The proposed solution makes use of Universally Unique Identifiers (UUID) version 4 variant 1 as defined in RFC 4122. UUIDs can be represented as 32 hexadecimal (base 16) digits separated by hyphens. UUIDs have 6 bits reserved for the predetermined variant and version bits, leaving 122 bits which can means a total of $5.3 \times 10^{36}$ combinations of UUID version 4 variant 1. A unique UUID is generated and stored as a token for later use.

### 3.2 Secure Hashing Algorithm (SHA)

Further, the proposed solution utilises the Secure Hashing Algorithm SHA-512 hashing algorithm which generates a 512 bit hash of 64 alphanumeric characters. In the proposed solution, the SHA-512 algorithm is used to hash the UUID generated during authentication and is stored for later use.

### 3.3 Implementation of the proposed solution

As per the 802.11 standards, the association process is followed by the authentication procedure according to Figure 1.1 below. The association process in both the AP and the client can be modified such that they verify the deauthentication frame. The modified association process (Figure 1.2) is described as follows:

- After the authentication procedure, the client randomly generates a UUID $u_1$. This UUID is used as a unique token and stored in memory. The token $u_1$ is then hashed using the SHA-512 hashing function. The hash $h_1$ is included in the Association Request Frame.

- When the AP receives the Association Request Frame, it checks whether the SHA-512 hash is already in its memory. This leads to either one of the following:

    - Case 1 - If the SHA-512 hash exists in memory of the AP, the AP randomly generates a UUID $u_2$ and stores it in memory. The token $u_2$ is then hashed using the SHA-512 hashing algorithm. The hash $h_2$ is included in the Association Response Frame and sent to the AP.

    - Case 2 - If the SHA-512 hash does not exist in the memory of the AP, the AP rejects the Association Request Frame considering it to be a replay attack.

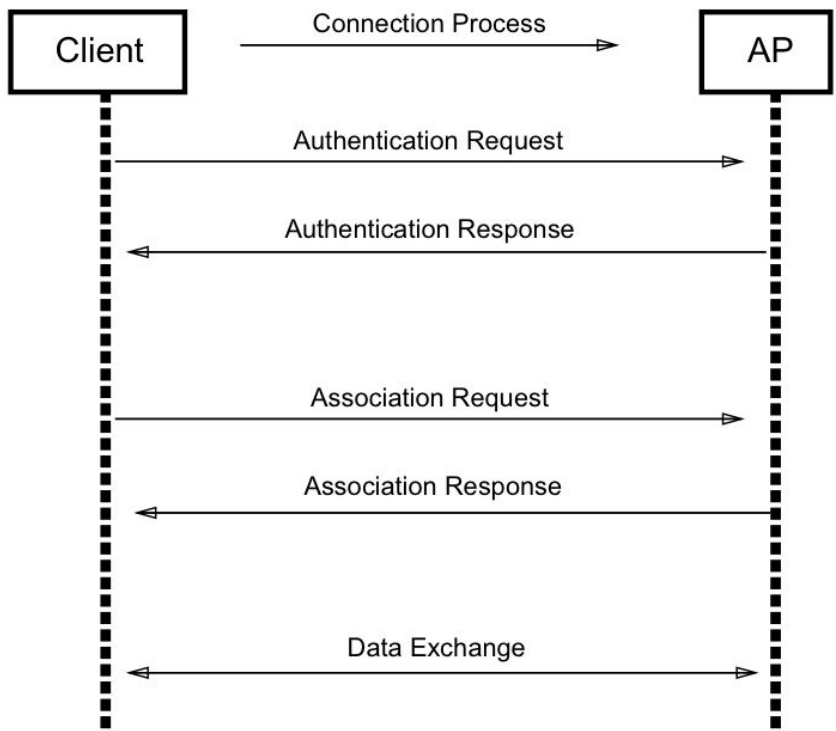

**Figure 1.1**
**Original Association Protocol**

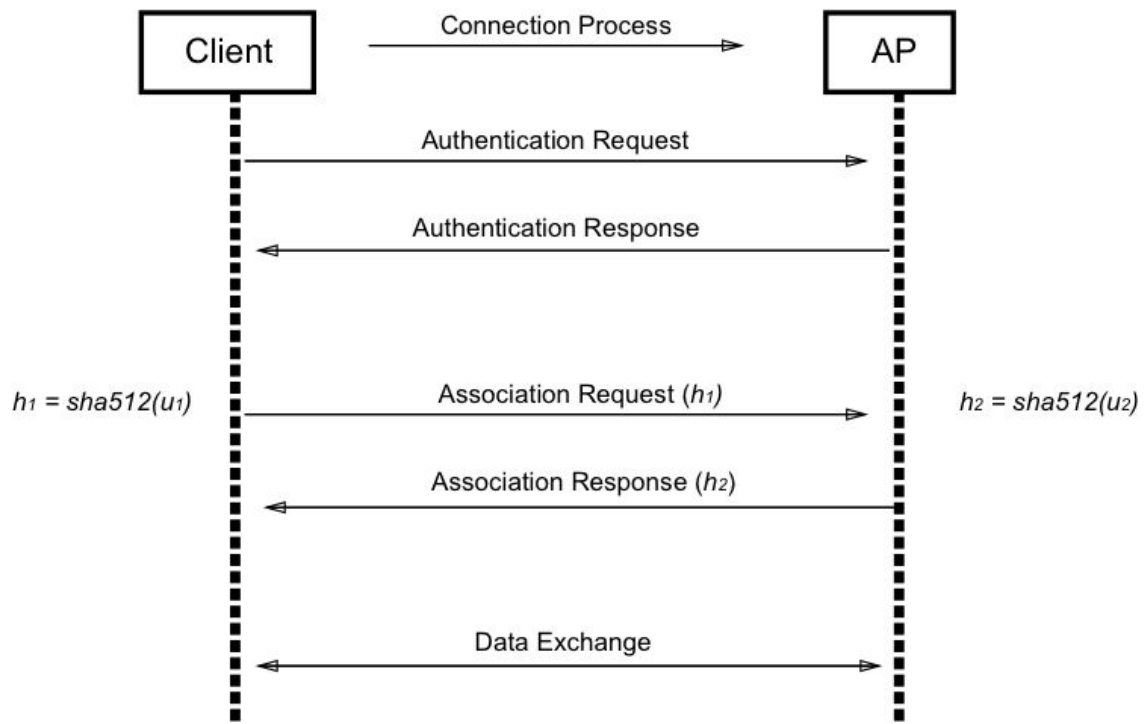

**Figure 1.2
Modified Association Protocol**

- Upon wanting to disconnect from the AP, the client will send the original token $u_1$ generated initially. Once this token is received by the AP along with the deauthentication request, the AP hashes this token using SHA-512. If the hash of the token received with the deauthentication request matches the SHA-512 hash originally sent during authentication, the deauthentication request is approved and the received token and its hash is deleted from memory. If the hash of the token sent along with the deauthentication request does not match the original SHA-512 hash sent during authentication, the deauthentication request is rejected.

- Likewise, when the AP wants to disconnect from the client, the AP sends the original token $U_2$ to the client in the deauthentication frame. When the client receives the token, it hashes the token using SHA-512 received with the deauthentication frame. If the hash of this token matches the SHA-512 hash received during authentication, the deauthentication request is approved and the received token and its hash is deleted from memory. If the hash received during the deauthentication process does not match the SHA-512 hash received during authentication, the deauthentication frame is rejected.

**Testing the new protocol**

According to the 802.11 standards, every deauthentication frame sent out by the AP or the client has a reason code (as shown in Table 1.1).

| Reason Code | Description |
| --- | --- |
| Code-0 | Reserved |
| Code-1 | Unspecified |
| Code-2 | prior authentication is not valid |
| Code-3 | station has left the basic service area or extended service area and is de-authenticated |
| Code-4 | Inactivity timer expired and station was disassociated |
| Code-5 | Disassociated due to insufficient resources at the access point |
| Code-6 | Incorrect frame type or subtype received from unauthenticated station |
| Code-7 | Incorrect frame type or subtype received from nonassociated station |
| Code-8 | Station has left the basic service area or extended service area and is disassociated |
| Code-9 | Association or reassociation requested before authentication is complete |
| Code-10 to 65535 | reserved |

Table 1.1
Reason Codes[5]

In order to define the working of the new protocol, each of the following situations associated with each reason code for disconnecting from the AP has been considered (Table 1.2) :

| Reason Code | Action |
| --- | --- |
| Code-0 | Reserved |
| Code-1 | This frame will be rejected. |
| Code-2 | Since the client is not already authenticated / associated with the AP, this frame will be ignored. |
| Code-3 | <ul><li>The client will simply send a disassociation frame with token $u_1$ included in the frame.</li><li>The AP compares the SHA-512 hash of the value sent included in the frame to the SHA-512 hash previously sent. If both the hash values match, then accept the deauthentication, else ignore the frame. While an attacker can spoof the MAC address of a client, the original token $u_1$ cannot be guessed due to $5.3 \times 10^{36}$ (5.3 undecillion) possibilities.</li></ul>(Incase the AP goes offline)<ul><li>The AP broadcasts the value for $u_2$ to clients respectively.</li><li>When the client receives this frame, it checks if the SHA-512 hash of the value received matches the initial SHA-512 hash exchanged during authentication. If the hashes match, then the client disassociates from the AP, else the frame is ignored.</li></ul> |
| Code-4 | Same as Code 3. |
| Code-5 | This will be the same as Reason Code 3. The AP broadcasts the value for $u_2$ to clients respectively. When the client receives this frame, it checks if the SHA-512 hash of the value received matches the initial SHA-512 hash exchanged during authentication. If the hashes match, then the client disassociates from the AP, else the frame is ignored. |
| Code-6 | Since the client is not already authenticated / associated with the AP, this frame will be ignored. |
| Code-7 | Since the client is not already authenticated / associated with the AP, this frame will be ignored. |
| Code-8 | <ul><li>The client will simply send a disassociation frame with token $u_1$ included in the frame.</li><li>The AP compares the SHA-512 hash of the value sent included in the frame to the SHA-512 hash previously sent. If both the hash values match, then accept the deauthentication, else ignore the frame. While an attacker can spoof the MAC address of a client, the original token $u_1$ cannot be guessed due to $5.3 \times 10^{36}$ (5.3 undecillion) possibilities.</li></ul> |
| Code-9 | Since the client is not already authenticated / associated with the AP, this frame will be ignored. |
| Code-10 to 65535 | reserved |

Table 1.2

## 4. Experiments

Since UUIDs have 5.3x1036 (5.3 undecillion) possibilities, it is practically impossible to use brute force to guess the UUID.

Microbenchmark tests can be performed to measure the time taken for generating UUIDs and using cryptographic hashing functions. Since computers and phones are significantly powerful, these microbenchmarking tests have been performed on IoT devices with low processing power and memory.

For this test the following devices have been made use of:

- ESP8266 WiFi Chip (L106 32-bit RISC microprocessor @ 80 MHz, 64 KiB of instruction RAM, 96 KiB of data RAM)
- Raspberry Pi 3 Model B (1.2 GHz 64/32-bit quad-core ARM Cortex-A53, 1 GB LPDDR2 RAM at 900 MHz)

| Task | Raspberry Pi (in seconds) | ESP8266 (in seconds) |
|---|---|---|
| Generating UUID | 0.076341 | 0.058025 |
| SHA-512 Hashing | 0.117223 | 0.123348 |
| Total Time | 0.193564 | 0.181373 |

The benchmarking results indicate that the proposed solution is fit for implementation on commercial devices.

## 4. Conclusion

This research paper proposes a solution applying cryptographic functions and generation of unique tokens to develop a secure association protocol that can prevent deauthentication attacks against 802.11 wireless networks. Furthermore, having been developed as an extension for 802.11 networks, this new association protocol also lends itself to extensive deployment to different APs and clients by simply performing a firmware upgrade.